\documentclass[aip,rsi,reprint,graphicx]{revtex4-1} 

\usepackage{graphicx}
\draft 

\begin{document}


\title{Towards hard X-ray imaging at GHz frame rate} 
\thanks{LA-UR-12-21057; Contributed paper submitted to the Proceedings of the 19th Topical Conference on High-Temperature Plasma Diagnostics, Monterey, California, May, 2012.\\}



\author{Zhehui Wang}
\email[]{zwang@lanl.gov}
\author{C. L. Morris}%
\author{J. S. Kapustinsky}
\author{K. Kwiatkowski}
\author{S.-N. Luo}

\affiliation{Los Alamos National Laboratory, Los Alamos, NM 87544}


\date{\today}

\begin{abstract}
Gigahertz (GHz) imaging using hard X-rays ($\gtrsim$ 10 keV) can be useful to high-temperature plasma experiments, as well as research using coherent photons from synchrotron radiation and X-ray free electron lasers. GHz framing rate can be achieved by using multiple cameras through multiplexing. The advantages and trade-offs of single-photon detection mode, when no more than one X-ray photon is detected per pixel, are given.  Two possible paths towards X-ray imaging at GHz frame rates using a single camera are a.) Avalanche photodiode arrays of high-Z materials and b.) Microchannel plate photomultipliers in conjunction with materials with large indices of refraction.
\end{abstract}


\maketitle 

\section{Introduction}
In laser-driven implosion experiments, characteristic implosion speeds can reach 10$^4$ to 10$^5$ m/s. To obtain a spatial resolution of 1 $\mu$m or less with minimum motion blur, the exposure time of an image can not exceed 10 to 100 ps. To obtain multiple frames of images using a single camera within a single experiment, the inter-frame time should be as close to the exposure time as possible. In other words, at least 1 GHz frame rate would be needed. Under a different context, advances in synchrotrons and X-ray free electron lasers (XFEL) ushered in and will continue to drive the transition from static structure imaging to fast and ultrafast imaging of structure evolution with sub-ns temporal resolution,~\cite{Chapman:2006,Barty:2008} if corresponding improvements in response time of X-ray imaging technologies can be made.  Sub-ps pulses of intense and coherent photons can be delivered by synchrotron and XFEL; motion blur is significantly reduced compared with using a conventional quasi-continuous plasma source.

One of the challenges in GHz imaging is high data rate. Dynamic range of an image, which is the ratio of the maximum to the minimum intensity ($\sim$ the noise level), can exceed 10$^6$.~\cite{Nugent:2009} Until recently, the only known medium that exceeded this dynamic range was imaging plate. Latest CMOS sensors reached 30-bit depth, or a dynamic range of 10$^9$. Imaging plates, like X-ray films, require a readout device to retrieve the image and subsequently a readout time on the order of a minute, which is slow compared with CCD and CMOS digital devices. But for multi-rame GHz imaging, the present readout speeds of the CCD or CMOS are not even sufficient for real-time data transfer. For a typical image with 1k$\times$1k array of pixels each with a dynamic range of 10$^6$ (20 bit), up to 2$\times$10$^{16}$ bits of data are generated per sec at 1 GHz frame rate. A few orders of magnitude improvement is still needed over the latest data transfer rate ($>$ 1 TB/s) through optical transmission. Data compression is one way to reduce the data transmission rate requirement. Meanwhile, it may be simply a matter of time before the required rate can be surpassed by advances in communication.  Therefore, we focus our discussions on the ultrafast detection of hard X-rays.

\section{Multiplexing}
\begin{figure}[htbp] 
   \centering
   \includegraphics[width=3in]{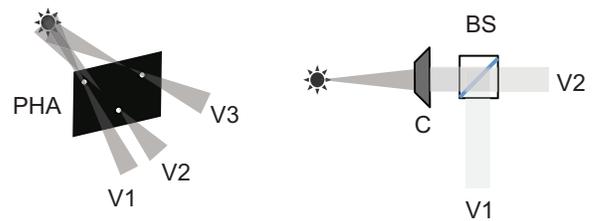} 
   \caption{Two possible multiplexing schemes to achieve GHz X-ray imaging rates using existing imaging devices, such as CCD and CMOS cameras. On the left, a pin-hole array (PHA) divides an objective view into `viewlets' (V1, V2, V3, etc). On the right, a single view is divided multiple times. A converter (C) is used to convert hard X-rays into, for example, visible light. Beam spliting (BS) for visible light is much easier to accomplish than for hard X-rays.}
   \label{fig:1x}
\end{figure}

At the present, there are two approaches to obtain multi-frame X-ray images with sub-ns resolution. One is X-ray streak camera.~\cite{Murnane:1990} The other is X-ray framing camera through fast gated microchannel plate (MCP).~\cite{Kilkenny:1991,Ortel:2006} A streak camera can achieve femtosecond temporal resolution, however, a single streak camera does not give a truly two dimensional (2D) spatial view of the object, since one of the dimensions of a 2D detector has to be devoted to time evolution. X-ray framing cameras rely on multiple pin-hole views of an object, see Fig.~\ref{fig:1x}. Only one frame is obtained for each `viewlet'. In particular, X-ray framing camera will have limited  use when there are not sufficient  photons.

Multiple-frame CCD and CMOS cameras do exist but are not fast enough. In the second generation proton radiography (pRad) imager, up to ten frames can be stored, with an interframe time of 150 ns. The pRad imager has a hybrid architecture, and it operates in a burst mode.  The imager uses 100 $\mu$m thick silicon sensor, which is bump bonded to a pixelated CMOS read-out integrated chip (ROIC).  

Through multiplexing, it is possible to combine the existing fast gating of MCP detection with improved multiple-frame CMOS or CCD cameras to achieve GHz frame rate. For example, by dividing the view of the object into 10 with one camera for each view, and if the camera can be improved to an interframe time of no more 10 ns,~\cite{Kleinfelder:2009} then by offsetting the camera start time by 1 ns, equivalent of 1 GHz rate is obtained. Besides further improvement in response time, this method relies on a very intense source of light to operate, which may not always be available. Below, we discuss how to obtain GHz framing rate with a single 2D camera.

\section{Single-photon detection mode}
\begin{figure}[htbp] 
   \centering
   \includegraphics[width=3in]{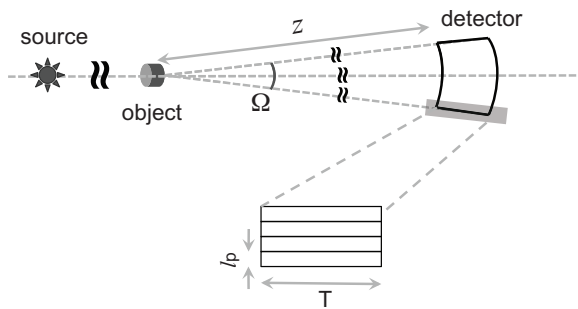} 
   \caption{A X-ray detector in ultrafast imaging should maintain a constant distance among different parts of the detector to the object. The length of each pixel (T, along the line of sight), determined by X-ray detection efficiency, is much greater than the lateral dimensions ($l_p$), which is determined by the requirement that no more than one photon impinges on each pixel.}
   \label{fig:1}
\end{figure}

We emphasize the detection process here, and do not elaborate on the interactions between the object and the photon field, nor the propagation of the photons to the detector. Both processes can cause image blur due to overlapping of signals from one pulse to the next in ultrafast imaging of mesoscale objects ($\sim$ 100 $\mu$m in size). For example, an photon can only travel 300 $\mu$m inside an object with a refractive index of one in 1 ps. Therefore for thicker objects, image blur due to overlaps of photons from pulse to pulse can potentially happen when different photons spend different amount of time inside the object. Similarly, a photon travels 0.03 m in 100 ps, the propagation times and the corresponding propagation distances to reach different parts of an area detector can not exceed these values. Ideally, a detector may be curved, as in Fig.~\ref{fig:1}, so that the distances from different parts of the detector to the object are the same. Otherwise, time corrections have to be made across the detector.

An ultimate detection scheme is `single-photon' detection (SPD) mode when each photon scattered or emitted from the object is detected separately. One simple approach to SPD is to place the detector as much away from the object as possible, so that the spatial density of the photons is less than one per pixel of an area detector. Therefore, the detector should have single-photon sensitivity and the number of the pixel exceeds the number of photons expected. No other detection mode exceeds the information content of a SPD mode. Another advantage of the SPD mode is that the dynamic range of each pixel can be quite small. One bit is in principle sufficient, and `0' stands for no photon hit and `1' for a photon hit. Each image therefore becomes a 2D map of 1's and 0's. Third, single-bit data allows the fastest analog-to-digital conversion, and can even eliminate the need for time stamping of each pulse, as long as all the pixels are synchronized. Fourth, the maximum number of frames of images can be stored on board before transmission for certain memory size. For example, the same storage for the pRad 10-frame imager (12 bit) mentioned above can store 120 frames of images in the SPD mode.

A trade-off of SPD is the large number of pixels and correspondingly large detector size, similar to the silicon microstrip detectors used in higher energy particle physics. In coherent diffractive imaging, the number of photons needed scales with resolution ($\delta$) as $\delta^{-3}$ or $\delta^{-4}$.~\cite{Shen:2004,Howells:2009} For an image formed by 10$^9$ photons and the ratio of the pixel number to the photon number of 10:1, the number of pixels per dimension is 10$^5$. Assuming a pixel size (also known as pitch) of 10 $\mu$m, the total area of the detector will be 1 m$^2$. In diffraction-limited imaging, the total detector area is equal to $z^2\lambda^2/\delta^2$ within a geometrical factor of order one. Here $z$ is the distance between the detector surface and the object, Fig.~\ref{fig:1}, $\lambda$ the wavelength and $\delta$ the resolution. Therefore, for fixed resolution and wavelength, the area of the detector increases with the detection distance as $z^{2}$.

There is a strong dependence of material thickness on photon energy for certain detection efficiency, in particular in the hard X-ray regime of energy. To detect an X-ray photon with an efficiency above 50\%, it is thus important to use high-Z materials to minimize the thickness of each pixel. X-ray absorption probability increases with the thickness ($T$) of the material as ($1-e^{-\mu\rho T}$), with $\mu$ being the mass attenuation coefficient, $\rho$ the mass density. At the efficiency of 50\%, T(50\%)$ = \ln 2/\mu\rho$. To detect 10 keV X-ray with silicon ($Z$=14), T(50\%) = 90 $\mu$m. For 50 keV X-rays, T(50\%)=12.9 mm. Using gallium arsenide (GaAs, $\bar{Z}$ = 32) for X-ray capture, T(50\%) = 36 $\mu$m and 424 $\mu$m for 10 keV and 50 keV respectively. Using indium antimonide (InSb, $\bar{Z}$ = 50) for X-ray capture, T(50\%) = 9 $\mu$m and 116 $\mu$m for 10 keV and 50 keV respectively. Cadmium telluride (CdTe, $\bar{Z} =$ 50) has the same average $Z$ as the InSb, therefore comparable thickness requirements for X-ray efficiency.

The lateral dimensions of each pixel, $l_p$ in Fig.~\ref{fig:1}, do not have to be as large as the thickness to better accommodate SPD mode.  The distance from the object to the detector $z \gg T$, the detector thickness. Therefore, the incoming X-rays are perpendicular to the surface of the pixel to within $T/z <$ 0.1\%. The possibility of edge effects, when an X-ray photon enters one pixel and ends up being detected in one of the neighboring pixels, is negligible, even for $l_p$ in the nm-range. Applications of deep sub-micron technology can make pixel size below 1 $\mu$m, smaller than the resolution of X-ray films. 

\section{GHz frame rate}
\begin{figure}[htbp] 
   \centering
   \includegraphics[width=3in]{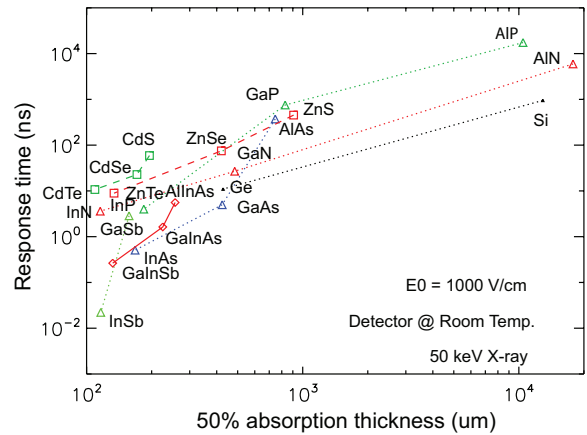} 
   \caption{Estimated response time based on electron drift velocities in different semiconductors for 50 keV X-ray. The detector is assumed to be at room temperature. The bias corresponds to an electric field of 10$^5$ V/m.}
   \label{fig:scin0x}
\end{figure}

Frame rate is limited to $1/\tau$, with $\tau$ being the total time required for X-ray capture, charge (or photon) transport, and data recording. Data transmission can add to $\tau$, and therefore should be avoided if possible. To achieve an imaging frame rate of at least a few GHz, the detector time response should be in the ps-range, which rules out a lot of common methods that use scintillators~\cite{zhu:2011} and semiconductors (such as CdTe) for X-ray capture. It was reported that 100 ps time resolution can be achieved using LaBr$_3$:Ce in conjunction with a silicon photomultipler (SiPM).~\cite{Schaart:2010} LaBr$_3$:Ce has a scintillating light decay time constant of 20 ns. There is no conflict here, since the time resolution of a time-of-flight type of detection, as in~\cite{Schaart:2010}, depends on fast digitizers (Acqiris DC282 or Agilent U1065A with a 8 GS/s rate, Tektronix DPO/DSA/MSO70000C reached 100 GS/s) with very small jitter of $\sim$ 1 ps. In imaging detectors through scintillating light, the decay time of the scintillator dominates the time response. Even one of fastest scintillators, BaF$_2$, has a fast decay constant about 600 ps, which is marginal for ultrafast imaging applications.  Therefore using scintillators is unlikely to deliver GHz rate of frame rate. Only two types of detectors reached single photon sensitivity and tens of ps response time: multi-channel plate photomultiplier tubes (MCP-PMTs) and single photon avalanche photo diodes (APDs).~\cite{Lacaita:1995} SiPM is a type of APD. Single photon sensitivity allows the highest possible efficiency in X-ray detection. The common factor in MCP-PMT and APD that allowed ps time response is the ballistic motion of electrons, corresponding to tens of eV to more than a few keV of kinetic energy and x10 to x100 faster than thermal drift velocities of electrons in semiconductors (0.1 to 1 eV in energy). In Fig.~\ref{fig:scin0x}, we have plotted the expected time response due to electron drift as a function of absorption layer thickness (50\% efficiency for 50 keV X-rays). The response time can be further reduced if the materials are cooled. Fig.~\ref{fig:scin0x} indicates that some of the fastest semiconductors can achieve sub-ns response time with 50\% detection efficiency, although further response-time reduction is needed for most of these materials.

Although the existing semiconducting X-ray detectors can not be used directly to achieve the GHz rate for X-ray imaging, these materials are worthy of further development along the line of APD arrays. Since X-ray absorption cross section increases with the atomic number $Z$, it is desirable to build APD with as large average $Z$ as possible. Using silicon as the reference, high-Z APD, such as germanium (Ge) APD, indium antimonide (InSb) APD, already exist.~\cite{Baertsch:1967,Assefa:2010} Advantage of the APD based imaging device are that they can be made into pixelated arrays and they can be readily integrated with the recording and readout microelectronics. 

The second option is based on MCP-PMTs. These devices have demonstrated faster response time than APD's but they need an ultrafast source of light. Cherenkov light~\cite{Afanasiev:2010} is the natural choice in this regard. To generate Cherenkov light from 10 keV electron requires an index of refraction ($n$) of not less than 5.1. So far no material is known to have this large value of refractive index. For 40 keV photoelectrons, $n \ge 2.7$. This is still very large but some materials are known to exceed this threshold.  Rutile (TiO$_2$) has $ 2.6 < n < 2.9$, silicon has $n =$ 3.96 at 590 nm, germanium has $n=4.0$, gallium arsenide has $n = 3.9$, gallium phosphide has $n=$ 3.5, lead sulfide has a peak $n$ = 3.9 near 340 nm. Unfortunately, most of these materials are not transparent, therefore further material search and development will be needed. In short, MCP-PMT based methods may be most useful for ultrafast imaging using very hard X-rays ($\gtrsim$ 40 keV) if certain Cherenkov light emitting material can be found.



%
%

%

\begin{acknowledgments}
 We thank Dr. Cris W. Barnes for very fruitful discussions and encouragement to carry out the work.
\end{acknowledgments}


\end{document}